\documentclass[prb,superscriptaddress,showpacs,aps]{revtex4}% Physical Review B

\usepackage{color}
\usepackage{graphicx}% Include figure files
\usepackage{dcolumn}% Align table columns on decimal point
\usepackage{bm}% bold math

\begin{document}

%\preprint{APS/123-QED}

\title{{Self consistent hydrodynamic description of the {plasma} wake field excitation induced by a relativistic charged-particle beam in an unmagnetized plasma}}
\author{Du\v san Jovanovi\'c}
\email{dusan.jovanovic@ipb.ac.rs} \affiliation{Institute of Physics, University of Belgrade, Pregrevica 118, 11080 Belgrade (Zemun), Serbia}
\author{Renato Fedele}
\email{renato.fedele@na.infn.it} \affiliation{Dipartimento di Fisica, Universit\`{a} di Napoli "Federico II" Complesso Universitario
M.S. Angelo,
%via Cintia, I-80126
Napoli, Italy}
\author{{Sergio De Nicola}}
\email{sergio.denicola@spin.cnr.it} \affiliation{SPIN-CNR, Complesso Universitario di M.S. Angelo, Napoli, Italy}
\affiliation{Dipartimento di Fisica, Universit\`{a} di Napoli "Federico II" Complesso Universitario
M.S. Angelo,
%via Cintia, I-80126
Napoli, Italy}
\affiliation{INFN Sezione di Napoli, Complesso Universitario di M.S. Angelo, Napoli, Italy}
\author{Tamina Akhter}
\email{tahminaphys@gmail.com} \affiliation{Dipartimento di Fisica, Universit\`{a} di Napoli "Federico II" Complesso Universitario
M.S. Angelo,
%via Cintia, I-80126
Napoli, Italy}
\affiliation{INFN Sezione di Napoli, Complesso Universitario di M.S. Angelo, Napoli, Italy}
\date{\today}
\author{Milivoj Beli\'c}
\email{milivoj.belic@qatar.tamu.edu} \affiliation{Texas A\&M University at Qatar, P.O. Box 23874 Doha, Qatar}

\begin{abstract}
A self-consistent nonlinear hydrodynamic theory is presented of the propagation of a long and thin relativistic electron beam, for a typical plasma wake field acceleration configuration in an unmagnetized and overdense plasma. The random component of the trajectories of the beam particles as well as of their velocity spread is modelled by an anisotropic temperature, allowing the beam dynamics to be approximated as a 3-D adiabatic expansion/compression. {It is shown that even in the absence of the {nonlinear plasma wake force}, the localization of the beam in the transverse direction can be achieved owing to the} nonlinearity associated with the adiabatic compression/rarefaction and a coherent stationary state is constructed. Numerical calculations reveal the possibility of the beam focussing and defocussing, but the lifetime of the beam can be significantly extended by the appropriate adjustments, so that transverse oscillations are observed, {similar to those} predicted within the thermal wave and Vlasov kinetic models.
\end{abstract}

\maketitle

\section{Introduction}\label{Introduction}

The concept of plasma-based particle accelerators, capable of accelerating charged particles to extremely high energies (beyond 10 TeV) has been put forward more than three decades ago and massive theoretical, computational and experimental efforts have been made in this field since then. One of the proposed schemes, often referred to as the plasma wake field accelerator (PWFA) consists in the injection of a relativistic electron (or positron) beam into the plasma, which creates a vacancy in the electron population of the background plasma while the plasma ions remain in place due to their large mass. After the passage of the beam{, usually referred to as the driving beam or simply the driver,} {in the wake of the latter}, plasma electrons rush in to restore the quasineutrality which produces a very large amplitude electrostatic plasma wave that propagates with the same velocity as the beam that has created it. The longitudinal electric field of such plasma wave can be used to accelerate particles to the energies that are higher that that of the driving beam. When the latter is properly bunched{, and the beam density is of the order of $10^{17}-10^{18}\, {\rm cm}^{-3}$,} an acceleration gradient {ranging from $0.01$ to $1$ GeV/cm} can be achieved, yielding the energy of the accelerated particles to be $3-4$ times bigger that that of the driving beam \cite{1985PhRvL..54..693C}. The dynamics of the driving beam in the direction transverse to its propagation is important not only for the feasibility of such scheme, but also for the quality ({i.e., lower emittance and higher brightness}) of the relativistic particle bunch generated in it. A high-intensity electron beam of finite extent can be subjected longitudinally to a two-stream instability, while in the transverse direction it can undergo self-focusing or defocusing, filamentation, and self-pinching. As for the quality of the bunch produced {this way}, transverse PWF can be used for its focussing {which contrasts the spreading due to the thermal emittance of the bunch. However, more generally, the concomitant presence of both {the} transverse PWF and thermal spreading of the bunch can be adequately used for the bunch manipulation. In particular, for sufficiently long bunches the PWF acts on the driving bunch itself, leading to the self consistent PWF excitation ({the} feedback of the wake field on the driver). Therefore, in such a physical {situation} the concomitance of transverse PWF and thermal spreading leads to the beam self modulation. On longer time scales, under suitable conditions, the latter can evolve in an unstable way (self modulation instability)}.

In the early theoretical works  \cite{1985PhRvL..54..693C} the feedback of the wake field on the driver and the resulting spatial evolution of the beam were not considered. Later, a selfconsistent theory based on the thermal wave model (TWM) \cite{1992PhRvA..45.4045F} was presented and applied to the study of the interaction between the plasma wake field and the driving relativistic electron beam, in an unmagnetized overdense plasma (whose density exceeds that of the beam) and in the case of a long beam ($L_\Vert\gg L_\bot$, where $L_\Vert$ and $L_\bot$ are its longitudinal and transverse characteristic lengths, respectively). Such model successfully described the beam's self-focusing and self-pinching equilibria, when the spot size of the beam was larger and smaller than the wavelength of the wake field, respectively. {More recently, this approach has been extended to the case of self consistent plasma wake field (PWF) excitation in a cold magnetized plasma for both long nonlaminar warm relativistic driving beam (quantum-like domain of TWM)
\cite{Fedele_et_al.2011,Tanjia_et_al.2011}
and relatively cold relativistic quantum driving beam (quantum domain)
%\cite{Fedele et al.PoP2012,Jovanovic et al.EPL2012}.
\cite{2012PhPl...19j2106F,2012EL....10055002J}.}
Similarly, the self-modulation of a relativistic charged-particle beam as {\textit{thermal matter wave envelope}} and the possibility of its destabilization was studied in Ref. \cite{2014JPhCS.482a2014F} using the quantum-like description provided by the {TWM} in which the beam dynamics is governed by a Zakharov-type system of equations, comprising a Poisson-like equation for the wake potential and a nonlinear Schr\"{o}dinger equation governing the spatiotemporal evolution of the thermal matter wave envelope, whose dispersion coefficient is proportional to the beam thermal emittance. In the strongly nonlocal regime, an Ermakov-Pinney type equation
%(also called the Sacherer's envelope equation)
for the evolution of the beam's cross-section was derived, and the possibilities were discussed of an instability leading to the beam blowup and of a stable self-modulation in the form of sausage-like transverse oscillations. {Fully similar results have been obtained by using a kinetic approach, where the study of the self consistent PWF excitation has been provided by the Vlasov-Poisson-type system of coupled equations
%\cite{Fedele et al. EPJDa, Fedele et al.EPJDb}.
\cite{2014EPJD...68..210F,2014EPJD...68..271F}.
In addition, with the same kinetic approach, the self-modulation instability has been successfully described in recent works
%\cite{T. Akhter et al.NIMA2016}}
\cite{2016NIMPA.829..426A}.}

In the present paper we derive a hydrodynamic description for the interaction between a long relativistic electron beam with an unmagnetized, overdense plasma. {Similarly} to the thermal wave model, the random component of the transverse motion of the beam particles is modelled by the finite temperature of the fluid, which is taken to be anisotropic. We demonstrate that in the case of a small but finite variation in the direction of the propagation, a nonlinearity associated with the adiabatic thermodynamic evolution of the beam produces the localization of the latter. Such coherent structure is in the state of an unstable equilibrium that, upon a small perturbation, may either collapse or {disperse} when the number of beam particles is, respectively, bigger or smaller than that in the stationary state. However, a fine tuning of the initial beam profile may sufficiently postpone these instabilities until other processes, not included in our model, enter the picture and determine the actual evolution of the beam. {However, for a thermodynamic process that is predominantly 2-D the related nonlinearity vanishes and, on the long timescale, the system is governed by the {nonlinear plasma wake force (given by the gradient of the effective potential)} that is responsible also for the betatron-like oscillations observed in the earlier works \cite{2014EPJD...68..210F,2014EPJD...68..271F,2014JPhCS.482a2014F,2016NIMPA.829..426A}.}

\section{Fluid and Vlasov descriptions}\label{relativisticVlasov}

The system {under study} consists of a quiescent plasma pierced by a relativistic beam of electrons, denoted in the rest of the paper by the subscripts $p$ and $b$, respectively. The beam propagates along the $z$ axis  with a relativistic speed $u$, $c-u\ll c$ and it is relatively long, $L_{b\Vert}\gg L_{b\bot}$, where $L_{b\Vert}$ and $L_{b\bot}$ are the characteristic lengths of inhomogeneity of the beam in the directions parallel and perpendicular to its direction of propagation, respectively. Expressing the electric and magnetic field in terms of the electrostatic and vector potentials, $\vec E = -\nabla\phi - \partial\vec A/\partial t$ and $\vec B=-\nabla\times\vec A$ and using the Coulomb gauge $\nabla\cdot\vec A=0$, the Poisson's equation and the Ampere's law take the familiar form
\begin{eqnarray}
&& \label{Poissons}
\nabla^2\phi = \left(e/\varepsilon_0\right)\left(n_b + n_p - n_i\right),
\\
&& \label{Amperes}
\left({\partial^2}/{\partial t^2}-c^2\nabla^2\right)\vec A = -\nabla\;\partial\phi/\partial t-\left({e}/{\varepsilon_0}\right)\left(n_b \vec v_b + n_p \vec v_p\right),
\end{eqnarray}
where $-e$ is the electron charge, $n_b$, $n_p$, and $n_i$ are the densities of the beam and plasma electrons, and of the plasma ions, $\vec v_b$, $\vec v_p$ are the hydrodynamic velocities of the beam and plasma electrons, while the temporal variation is considered as sufficiently rapid so that the plasma ions are immobile.
We restrict our study to regimes in which the temporal variation in the reference frame comoving with the beam is sufficiently slow and we conveniently introduce the variables
\begin{equation}\label{variables}
\vec{r}\,' = \vec{r} - \vec{e}_z \, u \, t, \quad \quad t'= t,\quad \quad \vec v_b\!^\prime = \vec v_b - \vec e_z u,
\end{equation}
which implies ${\partial}/{\partial t} = {\partial}/{\partial t'} - u \; {\partial}/{\partial z'}$. Furthermore, we assume that the electric field is sufficiently weak so that the fluid velocities $\vec v_p$ and $\vec v_b\!^\prime$ are nonrelativistic, i.e. we adopt the following scaling
\begin{equation}\label{nonrelativistic_fluid}
\partial/\partial t'\sim \vec v_b\!^\prime\cdot\nabla \sim \vec v_p\cdot\nabla \sim\epsilon \, u\;\partial/\partial z,   \quad {\rm where} \quad \epsilon\ll 1.
\end{equation}
The plasma density $n_p$ and fluid velocity $\vec v_p$ are readily calculated from the appropriate hydrodynamic equations. We consider a regime in which the density of the beam scales as $n_b\sim n_p-n_i \sim n_i \, v_p/c$ and, as a consequence, the Lorentz forces acting on an electron in the beam does not exceed the Coulomb force, $|\vec e_z u \times \vec B_0|\lesssim |\vec E|$, provided $u\sim c$.
We take also that the plasma electrons are nonrelativistic and that the potential $\phi$ is sufficiently small,
\begin{equation}\label{scaling_phi}
%e\nabla_\bot\phi \lesssim \epsilon \, m_0 u^2\,\partial/\partial z,\quad\quad
{e\phi/m_0 u^2\lesssim \epsilon \, (\partial/\partial z)/\nabla_\bot,}
\end{equation}
$m_0$ being the electron rest mass, which permits us to neglect all convective nonlinearities and also the Lorentz force in the momentum equation for the plasma electrons. We study a regime in which the background plasma is cold, and the continuity and momentum equations for plasma electrons take the simple form
\begin{eqnarray}
&&\label{contin_p}
-u\;{\partial n_p}/{\partial z} + n_{p 0} \nabla\cdot\vec v_p = 0,
\\
&&\label{momentum_p}
-u\;{\partial \vec v_p}/{\partial z} = \left(e/m_0\right)\left(\nabla\phi - u \; \partial\vec A/\partial z\right),
\end{eqnarray}
$n_{p 0}$ being the unperturbed density of plasma electrons. Eliminating $\vec v_p$ and using the Poisson's equation (\ref{Poissons}), we have
\begin{equation}\label{first_p}
\left(\omega_{pp}^2 + u^2\,\partial^2/\partial z^2\right)\nabla^2\phi = \left(e/\varepsilon_0\right)u^2\,\partial^2 n_b/\partial z^2,
\end{equation}
where $\omega_{pp} = \sqrt{n_{p 0} e^2/m_0\varepsilon_0}$ is the plasma frequency of the background plasma. With the same accuracy, from $z$-components of the momentum equation (\ref{momentum_p}) yielding ${v_p}_z=(e/m_0 u)(\phi - u A_z)$, and of the Ampere's law (\ref{Amperes}) we get
\begin{equation}\label{second_p}
\left(\omega_{pp}^2 + u^2\,\partial^2/\partial z^2- c^2\nabla^2\right)\left(\phi-u A_z\right)+ c^2\nabla^2 \phi = \left(e/\varepsilon_0\right)u^2\, n_b.
\end{equation}
Finally, combining Eqs. (\ref{first_p}) and (\ref{second_p}) we obtain
\begin{equation}\label{final_p}
\left(\omega_{pp}^2 + u^2\,\partial^2/\partial z^2\right)\left(\omega_{pp}^2 + u^2\,\partial^2/\partial z^2 - c^2\nabla^2\right) \left(\phi - u A_z\right) = \left(e/\varepsilon_0\right)\left[\omega_{pp}^2 + \left(u^2-c^2\right)\partial^2/\partial z^2\right]u^2 n_b,
\end{equation}
which, for a long beam, $\nabla_\bot\gg\partial/\partial z$, whose velocity is relativistic, $u\approx c$, simplifies to the well known result
\begin{equation}\label{final_p_apr}
\left(\omega_{pp}^2 - c^2\nabla_\bot^2\right) \left(\phi - u A_z\right) = \left(e/\varepsilon_0\right)u^2 n_b,
\end{equation}
%or
%\begin{equation}\label{final_p_apra}
%\left[1 - \left(c^2/\omega_{pp}^2\right)\nabla_\bot^2\right] \left(\phi - u A_z\right) = -\left(m_0 u^2/e\right) \left(n_b/n_{p0}\right),
%\end{equation}
The density of the beam, $n_b$, is calculated from the well-known textbook expression for the relativistic Vlasov equation,
\begin{equation}\label{rel_Vlasov}
\frac{\partial f_b}{\partial t} + \vec{\rm v}\cdot\frac{\partial f_b}{\partial\vec{r}} - e\left(\vec{E} + \vec{\rm v}\times\vec{B}\right)\cdot\frac{\partial f_b}{\partial\vec{\rm p}} = 0,
\end{equation}
where $f_b(t,\vec{r},\vec{\rm p})$ is the distribution function of the beam electrons, while $\vec{\rm v}$ and $\vec{\rm p}$ are the electrons' velocity and momentum, that are mutually related as
\begin{equation}\label{vel_momen}
\vec{\rm v} = \frac{\vec{\rm p}}{m_0\sqrt{1 + {\rm p}^2/\left(m_0^2 c^2\right)}} \quad \Leftrightarrow \quad
\vec{\rm p} = \frac{m_0\vec{\rm v}}{\sqrt{1 - {\rm v}^2/c^2}}.
\end{equation}
However, solving the full Vlasov equation can be very tedious in a general case. In this paper, we do not account for the kinetic effects, most notably the wave breaking of the Langmuir wave associated with the plasma wake and the resonant particle effects (such as the trapping and acceleration of resonant electrons), and we solve instead the appropriate moments of the Vlasov equation (\ref{rel_Vlasov}). Multiplying the Vlasov equation (\ref{rel_Vlasov}) by $1$, $\vec{\rm p}$, and ${\rm v}_i\, {\rm p}_j$, respectively, and integrating for the entire momentum space, we obtain the relativistic fluid equations of continuity and momentum, and the equation for the $(i,j)$ component of the pressure tensor
\begin{eqnarray}
&&\label{rel_cont}
\partial n_b/\partial t + \nabla\cdot\left(n_b \vec v_b\right)=0
\\
&&\label{rel_moment}
\left(\partial/\partial t + \vec v_b\cdot\nabla\right)\vec p_b = -e\left(\vec E + \vec v_b\times \vec B\right)-\left(\nabla\cdot \mathbf{P}\right)/n_b,
\\
&&\label{rel_pres_tens}
\partial P_{k,l}/\partial t + \nabla\cdot\left(P_{k,l}\,\vec v_b\right) + P_{j,k}\;\partial {v_b}_l/\partial x_j + P_{j,l}\;\partial {v_b}_k/\partial x_j = 0,
\end{eqnarray}
where we have used the standard notation from tensor algebra $\nabla\cdot\mathbf{P} = \vec e_\alpha \, \partial P_{\alpha, \beta}/\partial x_\beta$. In the equations (\ref{rel_pres_tens}) for the stress tensor in a collisionless plasma, we have neglected both the thermal convection and the off-diagonal terms of $\mathbf{P}$, i.e. we have $\mathbf{P} = P_\bot(\mathbf{I} - \vec{e}_z \vec{e}_z) + P_\Vert\,\vec{e}_z \vec{e}_z$ where $\mathbf{I}$ is a unit tensor, viz. $I_{\alpha, \beta} = \delta_{\alpha, \beta}$ and $\delta_{\alpha, \beta}$ is the Kronecker delta. We rewrite these equations using the variables defined in Eq. (\ref{variables}) and introduce also $\vec v_b\!^\prime = \vec v_b - u\,\vec e_z$. When the fluid velocity $\vec v_b\!^\prime$ is nonrelativistic, in the momentum equation (\ref{rel_moment}) we may set
\begin{equation}\label{effective_E}
\vec E + \vec v_b\times \vec B \; \approx \; \vec E + u \,\vec e_z \times \vec B \; \approx \; - \nabla\left(\phi - u A_z\right),
\end{equation}
and our fluid equations{, rewritten through the variables defined in Eq. (\ref{variables}),} readily obtain the form
\begin{eqnarray}
&&\label{rel_cont_1}
\partial n_b/\partial t' + \nabla\cdot\left(n_b \vec v_b\!^\prime\right)=0
\\
&&\label{rel_moment_1}
\left(\partial/\partial t' + \vec v_b\!^\prime\cdot\nabla\right)\vec p_b = e\nabla\left(\phi - u A_z\right) - \left(\nabla_\bot P_\bot + \vec e_z\;\partial P_\Vert/\partial z \right)/n_b,
\\
&&\label{rel_pres_tens_1}
\partial P_\bot/\partial t' + \nabla\cdot\left(P_\bot\,\vec v_b\!^\prime\right) + P_\bot\nabla_\bot\cdot\vec v_b\!^\prime = 0,
\\
&&\label{rel_pres_tens_2}
\partial P_\Vert/\partial t' + \nabla\cdot\left(P_\Vert\,\vec v_b\!^\prime\right) + 2 P_\Vert \; {\partial v'_b}_z/\partial z = 0.
%\partial P_\Vert/\partial t' + \nabla\cdot\left(P_\Vert\,\vec v_b\!^\prime\right) + 2 P_\Vert\; \vec{b}\cdot\left(\vec{b}\cdot\nabla\right)\vec v_b\!^\prime = 0,
\end{eqnarray}
%where $\vec{b}=\vec{B}/|\vec{B}|$ is a unit vector along the magnetic field.
The pressures $P_\bot$ and $P_\Vert$ will be calculated using the Grad's approximation, expressed through the equilibrium solution of the Vlasov equation and using a limited number of its moments. As the equilibrium solution we use the J\"{u}ttner distribution with a particle drift, which is one of the several relativistic versions of the Maxwellian distribution function that can be found in the literature. In the case of an anisotropic electron temperature, it has the form \cite{2013PhPl...20d4501N}:
\begin{equation}\label{Juttner}
f_0\left(\vec{\rm p}_\bot, {\rm p}_z\right) = C \, \exp\left[-\alpha_\bot \gamma_u\left(\gamma - \beta_u \, {\rm p}_z/mc\right)-\left(\alpha_\Vert-\alpha_\bot\right)\sqrt{1 + \gamma_u^2\left({\rm p}_z/m_0 c - \beta_u\gamma\right)^2}\right],
\end{equation}
where
\begin{equation}\label{relativistic_notation}
\alpha_\bot = m_0 c^2/T_{0\bot},  \quad  \alpha_\Vert = m_0 c^2/T_{0\Vert},  \quad  \gamma = \sqrt{1+(\vec{\rm p}_\bot^{\,2} + {\rm p}_z^2)/m_0^2 c^2}, \quad \gamma_u = 1/\sqrt{1-\beta_u^2},  \quad  \beta_u = u/c,
\end{equation}
and $C$ is determined from the normalization, $\int_{-\infty}^\infty d^3\vec{\rm p} \; f_0 = n_0$. $T_{0\bot}$ and $T_{0\Vert}$ are the perpendicular and parallel temperatures in the comoving frame \cite{1994tdcp.book.....R}.
% $C = \alpha \, [4 \pi \, m_0^2 \, c^3 \gamma_u \, K_2(\alpha)]^{-1}$
%$C = n_0 / \,[2 \pi \, m_0^2 \, c^3 \gamma_u \, K_2(\alpha)]$ and $K_2$ is the modified Bessel function of second order.

For nonrelativistic temperatures $\alpha_{\bot, \Vert}\gg 1$, the J\" uttner distribution (\ref{Juttner}) reduces to a shifted Maxwellian, with appropriate relativistic corrections to the electron mass and temperature
\begin{equation}\label{shifted_Maxwellian}
f_0\left(\vec{\rm p}_\bot, {\rm p}_z\right) = \frac{n_0}{\left(2\pi m_0\right)^\frac{3}{2}  T_{0\Vert}^\frac{1}{2} \, T_{0\bot} \,\gamma_u}\; \exp\left[-\frac{\vec{\rm p}_\bot^{\,2}}{2 m_0 T_{0\bot}} - \frac{\left({\rm p}_z - m_0 u \, \gamma_u\right)^2}{2 m_0 T_{0\Vert} \gamma_u^2}\right].
\end{equation}
A reliable approximation for the distribution function in the presence of finite potentials $\phi$ and $A_z$, referred to as the Grad's moment approximation, is obtained when in the equilibrium distribution (\ref{shifted_Maxwellian}) we make the substitutions $n_0\to n_b$, $\vec{\rm p}\to\vec{\rm p}-\vec p_b$, and $T_{0\bot,\Vert}\to T_{\bot,\Vert}$.
Perpendicular and parallel pressures $P_\bot$ and $P_\Vert$ are then readily calculated as the appropriate integrals of such distribution function, viz.
\begin{equation}\label{approx_pressures}
P_\bot = n_b T_\bot/\gamma_u \quad {\rm and} \quad P_{\Vert} = n_b T_\Vert \gamma_u.
\end{equation}
From these expressions for the perpendicular and perpendicular pressures we note that $P_{\Vert}/P_{\bot} \sim \gamma_u^2 \; T_{\Vert}/T_{\bot}$. This may be a very big number and permits us to estimate the relevant scalings of the beam dynamics in the following way:
\\
In the regime $n_b \, e (\phi-u A_z)\sim P_\bot\ll P_\Vert$, the parallel convection is negligible, viz. ${v_b}'_z\partial/\partial z \ll \vec v_b\!^\prime\cdot \nabla_\bot$, when
\begin{equation}\label{sc1}
\partial/\partial z \ll \left(T_\bot/T_\Vert\right)^\frac{1}{2}\gamma_u^{-1} \, \nabla_\bot
\end{equation}
Eqs. (\ref{rel_pres_tens_1}) and (\ref{rel_pres_tens_2}) for the pressures $P_\bot$ and $P_\Vert$, remain coupled even for a weak $z$-dependence, Eq. (\ref{sc1}), provided $P_\Vert \, \partial{v_b}'_z/\partial z \gtrsim P_\bot \nabla_\bot\cdot\vec v_b\!^\prime $, which corresponds to
\begin{equation}\label{sc2}
\partial/\partial z \gtrsim \left(T_\bot/T_\Vert\right)\gamma_u^{-2} \, \nabla_\bot
\end{equation}
Conversely, Eq. (\ref{rel_pres_tens_2}) for $P_\Vert$ is decoupled and can be discarded when $P_\Vert \, \partial{v_b}_z/\partial z \ll P_\bot \nabla_\bot\cdot \vec v_b\!^\prime $, which corresponds to
\begin{equation}\label{sc3}
\partial/\partial z \ll \left(T_\bot/T_\Vert\right)\gamma_u^{-2} \, \nabla_\bot.
\end{equation}
Finally, using $P_\Vert = \gamma_u^2 \,  P_\bot(T_\Vert/T_\bot)$ with $T_\Vert/T_\bot \approx {\rm constant}$, and making an appropriate combination of Eqs. (\ref{rel_pres_tens_1}), (\ref{rel_pres_tens_2}), and (\ref{rel_cont_1}), we obtain
\begin{equation}\label{adiabatic_process}
\left(\partial/\partial t' + \vec v_b\!^\prime\cdot\nabla\right)\left(n_b^{-\kappa} \, P_\bot\right) = 0,
\end{equation}
which is, essentially, the equation of state for an adiabatic thermodynamic process. Here, from the scalings (\ref{sc2}) and (\ref{sc3}), we note that we have $\kappa = 5/3$ for a not too long beam (\ref{sc2}), and $\kappa = 2$ for an extremely long beam (\ref{sc3}).

\section{Numerical results}

Now, under the scalings (\ref{sc1}) and (\ref{sc2}), or (\ref{sc3}), we can rewrite our basic equations (\ref{final_p}) and (\ref{rel_cont_1})-(\ref{rel_pres_tens_2}) as
\begin{eqnarray}
&& \label{final_p_apr_F}
\left[\left(c^2/\omega_{pp}^2\right)\nabla_\bot^2 - 1\right] U_w = -\left(m_0 u^2 /n_{p0}\right)\,n_b,
\\
&&\label{rel_cont_F}
\partial n_b/\partial t' + \nabla_\bot\cdot\left(n_b \, \vec {v_b}^\prime_\bot\right)=0,
\\
&&\label{rel_moment_1_F}
m_0 \gamma_u \left(\partial/\partial t' + \vec {v_b}^\prime_\bot\cdot\nabla_\bot\right)\vec {v_b}^\prime_\bot = \nabla_\bot\left(U_w - \tau \, n_b^{\kappa-1}\right),
\end{eqnarray}
where $U_w = e(\phi-u A_z)$, $\tau = (T_{0\bot}/n_{b 0}^{\kappa-1})[\kappa/(\kappa-1)]$, and $\kappa$ takes the values $5/3$ and $2$ under the scalings (\ref{sc2}) and (\ref{sc3}), respectively. Using scaled variables $t''=t'/{\rm T}$, $\vec{v_b}^{\prime\prime}_\bot = \vec {v_b}^\prime_\bot/{\rm V}$, $\vec{r}\;^{\prime\prime}\!\!\!\!\!_\bot =  \vec{r}_\bot/{\rm R}$, $U_w''=U_w/{\rm W}$, and $n_b''=n_b/{\rm N}$, where
\begin{equation}\label{normalizations}
{\rm N} = n_{bc}\left(\frac{n_{p0}}{n_{bc}}\;\frac{T_{bc}}{m_0 u^2}\;\frac{\kappa}{\kappa-1}\right)^\frac{1}{2-\kappa},\quad
{\rm W} = m_0 u^2\; \frac{\rm N}{n_{p0}}, \quad
{\rm V} = u \, \sqrt{\frac{\rm N}{\gamma_u n_{p0}}}, \quad
{\rm T} = \frac{c}{\omega_{pp} u}\,\sqrt{\frac{\gamma_u n_{p0}}{\rm N}}, \quad
{\rm R} = \frac{c}{\omega_{pp}},
\end{equation}
where $n_{bc}$ and $T_{bc}$ are the beam density and temperature in the centre of the structure, respectively, our equations (\ref{final_p_apr_F})-(\ref{rel_moment_1_F}) are rewritten in a dimensionless form an with no physical parameters involved, viz.
\begin{eqnarray}
&& \label{final_p_apr_F1}
\left(\nabla_\bot^2 - 1\right) U_w = -n_b,
\\
&&\label{rel_cont_F1}
\partial n_b/\partial t + \nabla_\bot\cdot\left(n_b \vec {\;v_b}_\bot\right)=0,
\\
&&\label{rel_moment_1_F1}
\left(\partial/\partial t + \!\vec {\;v_b}_\bot\cdot\nabla_\bot\right)\!\vec {\;v_b}_\bot = \nabla_\bot\left(U_w - n_b^{\kappa-1}\right),
\end{eqnarray}
where for simplicity, the notations $''$ are omitted hereafter. It is worth noting that in the derivation of the Poisson's-like equation (\ref{final_p_apr}) or (\ref{final_p_apr_F1}), which accounts for the plasma response to the beam density $n_b$, we have neglected small terms of the order $\partial/\partial t'$ and $\vec v_p\cdot\nabla$ (the latter are assumed to be of the same order as the convective terms $\vec v_b\!^\prime\cdot\nabla$ in the hydrodynamic equations for the beam). Thus, Eqs. (\ref{final_p_apr_F1})-(\ref{rel_moment_1_F1}) describe essentially an adiabatic evolution of the potential $U_w$ in which the plasma instantaneously responds to the slow changes of the beam density $n_b$.
A stationary solution (denoted with the subscript "$0$"), with $\partial/\partial t = \vec {v_b}_\bot = 0$ is readily obtained, substituting $n_{b 0} = U_{w 0}^\frac{1}{\kappa-1}$ into the Poisson's-like equation (\ref{final_p_apr_F1}), yielding an equation from which all physical parameters have been scaled out, viz.
\begin{equation}\label{hole-like}
\left(\nabla_\bot^2 - 1\right)U_{w 0} + U_{w 0}^\frac{1}{\kappa-1}=0.
\end{equation}
When the beam is extremely long, i.e. in the strictly 2D case (\ref{sc3}) with $\kappa=2$, the above Eq. (\ref{hole-like}) is linear {and can not possess a localized solution that is well-behaved at $r=0$. An appropriate localization in such strictly 2D regime is possible only in the presence of a finite fluid velocity, viz. $\vec {v_b}_\bot\ne 0$, when the convective nonlinearity associated with the curlfree component of $\vec {v_b}_\bot$ produces a {nonlinear plasma wake force} on the slow time scale. The latter drives the modulational instability which, eventually, saturates into a coherent soliton state described in Refs. \cite{2014EPJD...68..210F,2014EPJD...68..271F,2014JPhCS.482a2014F,2016NIMPA.829..426A}.}

\begin{figure}[htb]
\centering
\includegraphics[width=70mm]{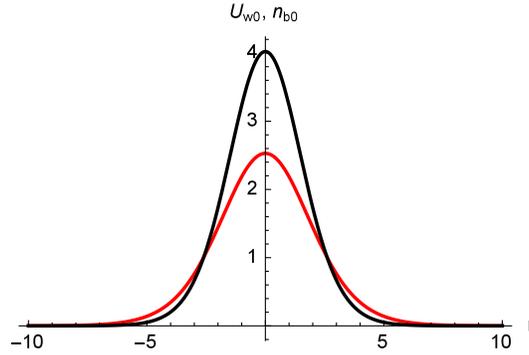}
\caption{Cylindrically symmetric stationary structure, found as the solution of Eq. (\ref{hole-like}) for $U_{w0}$ (red) and the corresponding $n_{b0}$ (black), in a three-dimensional case with $\kappa = 5/3$.}
\label{hole1}
\end{figure}

Conversely, in the presence of a finite dependence along $z$, (\ref{sc2}), our Eq. (\ref{hole-like}) contains a nonlinear term $U_{w 0}^{3/2}$ which produces the localization of the solution. Physically, the localization comes from the 3-D adiabatic process described by the appropriate equation of state (\ref{adiabatic_process}).
%The spatial shape of such nonlinear solution is universal and does not depend on the temperature-related quantity $\tau$.
The cylindrically symmetric solution of Eq. (\ref{hole-like}) is displayed in Fig. \ref{hole1}.

The stability of the adiabatic evolution of such solution with respect to small perturbations can be assessed if we linearize Eqs. (\ref{final_p_apr_F1})-(\ref{rel_moment_1_F1}) around the stationary solution, setting $n_b=n_{b 0} + \epsilon n_{b1}$, $\!\vec {\; v_b}_\bot = \epsilon \!\!\! \vec {\;\; v_{b1}}_\bot$, and $U_w = U_{w 0} + \epsilon U_{w 1}$, when to the first order in the (small) bookkeeping parameter $\epsilon$ we obtain
\begin{eqnarray}
&&\label{veza_U1_n1}
n_{b1} = -\left(\nabla_\bot^2 - 1\right) \, U_{w1},
\quad {\rm where} \quad
\nabla_\bot^2 = \frac{\partial^2}{\partial r^2} + \frac{1}{r}\frac{\partial}{\partial r}
\\
&&\label{perturbed}
-\omega^2 n_{b1} +\frac{1}{r}\frac{\partial}{\partial r}\left\{n_{b0}\, r \, \frac{\partial}{\partial r}\left[U_{w1}-\left(\kappa-1\right)n_{b0}^{\kappa - 2} n_{b1}\right]\right\} = 0.
\end{eqnarray}
Here we have applied the Fourier transformation in time and assumed a cylindrically symmetric perturbation. Multiplying Eq. (\ref{perturbed}) by $r \, [U_{w1}^*-(\kappa-1)\, n_{b0}^{\kappa - 2} n_{b1}^*]$, where $^*$ denotes the complex conjugate, doing some simple manipulations and taking that all quantities vanish at $r\to\infty$, we arrive at
\begin{equation}\label{omega_estimate}
\omega^2 = -\frac{\int_0^\infty{r\, dr \, n_{b0}\left|\left(\partial/\partial r\right)\left[U_{w1}-\left(\kappa-1\right)n_{b0}^{\kappa-2}n_{b1}\right]\right|^2}}
{\int_0^\infty{r\, dr \, \left[\left|U_{w1}\right|^2 + \left|\partial U_{w1}/\partial r\right|^2-\left(\kappa-1\right)n_{b0}^{\kappa-2}\left|n_{b1}\right|^2\right]}},
\end{equation}
Obviously, the right-hand-side of Eq. (\ref{omega_estimate}) is a real quantity and its numerator is positive definite, which implies that small perturbations
of the stationary solution are either purely growing/damped ($\omega^2 <0$) or purely oscillating functions of time ($\omega^2 >0$), depending on the sign of the {denominator}. Using the following simple estimate, we can indicate that both oscillating and growing/damped solutions may exist. If we estimate the stationary beam profile by a rectangular shape, $n_{b0}(r)\approx n_{bm}\, h(r_m-r)$ with $n_{bm}\approx 4$ and $r_m\approx 2.5$, we can rewrite Eq. (\ref{perturbed}) simply as
\begin{eqnarray}
&&\nonumber
\left(\nabla_\bot^2+k_+^2\right)\left(\nabla_\bot^2+k_-^2\right)\, U_{w1} = 0, \quad {\rm when} \quad r<r_m
\\
&&\label{rectangular_nb0}
\left(\nabla_\bot^2-1\right)\, U_{w1} = 0, \quad {\rm when} \quad r>r_m,
\end{eqnarray}
where
\begin{equation}\label{wavenumbers}
k_\pm^2 = \frac{1}{2\left(\kappa-1\right)n_{b m}^{\kappa-1}}\left\{\omega^2+n_{b m}-\left(\kappa-1\right)n_{b m}^{\kappa-1}\pm
\sqrt{\left[\omega^2+n_{b m}-\left(\kappa-1\right)n_{b m}^{\kappa-1}\right]^2 + 4 \, \omega^2\left(\kappa-1\right)n_{b m}^{\kappa-1}}\right\}.
\end{equation}
A small perturbation is now readily written as $U_{w1} = a_+ J_0(k_+ r) + a_- J_0(k_-r)$ for $r<r_m$ and $U_{w1} = b\, K_0(r)$ for $r>r_m$, where $a_\pm$ and $b$ are constants of integrations. From the requirements that $U_{w1}$, $\partial U_{w1}/\partial r$, and $\nabla_\bot^2 U_{w1}$ are continuous functions at the edge of the beam, $r=r_m$ (i.e. that the Coulomb and pressure forces on the electron fluid are finite and that there are no surface charges at the edge of the beam) and eliminating the constants of integration $a_\pm$ and $b$ we readily obtain the dispersion relation for the frequency of the linear mode $D(\omega^2)=0$, where
\begin{equation}\label{disprel}
D\left(\omega^2\right) = k_-\left(1+k_+^2\right)\frac{J_1\left(k_- r_m\right)}{J_0\left(k_-r_m\right)} +
k_+\left(1+k_-^2\right)\frac{J_1\left(k_+ r_m\right)}{J_0\left(k_+ r_m\right)} -
\left(k_-^2 - k_+^2\right)\frac{K_1\left(r_m\right)}{K_0\left(r_m\right)}.
\end{equation}

\begin{figure}[htb]
\includegraphics[width=70mm]{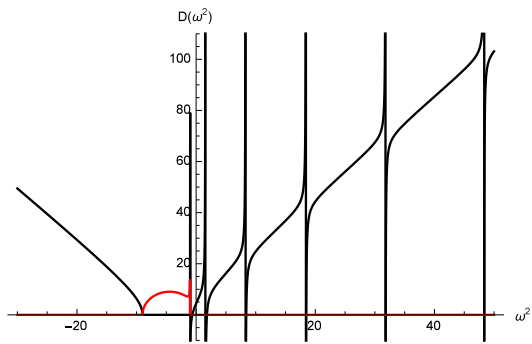}
\caption{Real (black) and imaginary (red) parts of the approximative dispersion function $D(\omega^2)$, defined in Eq. (\ref{disprel}). The equation $D(\omega^2)=0$ reveals the existence of two unstable modes, with $\omega^2<0$, and a sequence of oscillation modes with $\omega^2>0$.}  \label{disp_func}
\end{figure}
Dispersion function $D(\omega^2)$ is displayed in Fig. \ref{disp_func}. We can readily see that there exist two unstable modes featuring $\omega^2<0$, with the more unstable one corresponding to the branch point of the characteristic wavenumbers $k_\pm$, see Eq. (\ref{wavenumbers}). Besides these, owing to the periodicity of the Bessel functions $J_0$ and $J_1$, there is also a sequence of oscillating solutions with purely real frequencies, i.e. with $\omega^2>0$. A numerical solution of the dispersion relation $D(\omega^2) = 0$, using Eq. (\ref{wavenumbers}) with $n_{bm}= 4$ and $r_m= 2.5$ yields $\omega^2 = -9, \; -0.633, \; 1.820, \; 8.415, \; 18.49, \; 31.82, \; 48.36, \;...$ . Using the normalizations defined in Eq. (\ref{normalizations}) and $\kappa = 3/2$, we find that the effective width of the stationary state of the beam is $r_m \sim 2.5 \; c/\omega_{pp}$, while the characteristic frequencies and growthrates of its linear perturbations scale (in physical units) as $\omega \sim \omega_{pp} (n_{p0}/\gamma_u\, n_{b0})^\frac{1}{2}(T_{b0}/m_0 c^2)\sim \omega_{pp}/\gamma_u^2$.

\begin{figure}[htb]
\includegraphics[width=120mm]{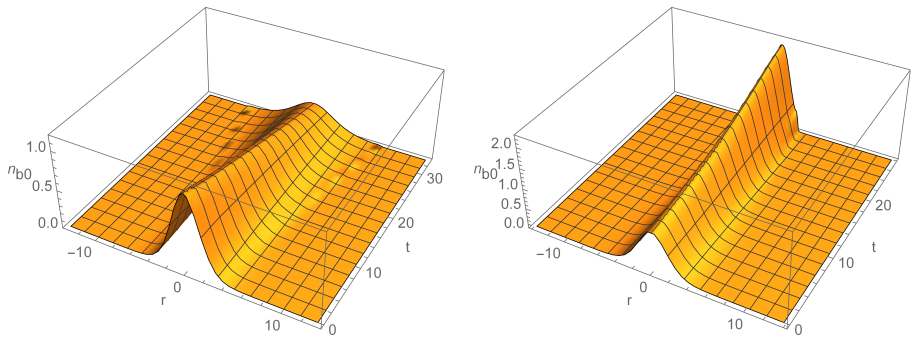}
\caption{The solutions of the full system of nonlinear equations (\ref{final_p_apr_F1})-(\ref{rel_moment_1_F1}). The initial conditions are adopted as $n_b(r, t=0) = n_{b0}(\alpha r)$ and $U_w(r, t=0) = U_{w0}(\alpha r)$, with $\alpha = 1.1$ (left) and $\alpha = 0.95$ (right).}  \label{DG}
\end{figure}

\begin{figure}[htb]
\includegraphics[width=120mm]{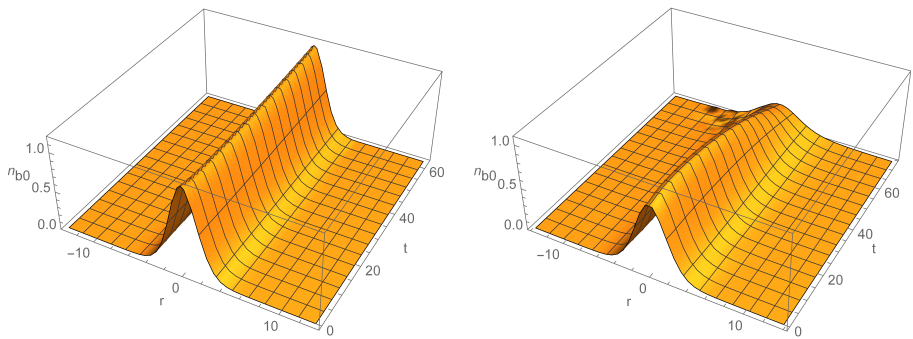}
\caption{Same as Fig. \ref{DG}, except that the initial conditions are adopted so as to contain the same number of electrons as a stationary beam, viz. $n_b(r, t=0) = \alpha^2 n_{b0}(\alpha r)$ and $U_w(r, t=0) = \alpha^{2(\kappa-1)} U_{w0}(\alpha r)$, with $\alpha = 1.05$ (left) and $\alpha = 0.9$ (right).}  \label{DG2}
\end{figure}

We have also solved numerically the full system of equations (\ref{final_p_apr_F1})-(\ref{rel_moment_1_F1}), see Figs. \ref{DG} and \ref{DG2}. {The solution appears to be spreading when the beam contains less electrons than the stationary state shown in Fig. \ref{hole1}, and collapsing when it contains more electron than the stationary state. Conversely, when we adopted an initial condition that contained the same number of electrons as the stationary state but with a different spatial profile, the solution remained stable for a considerably longer time, and even performed a limited number of oscillations (one or two) before the final collapse or spreading took place.} Such behavior comes from the fact that, with our somewhat arbitrary choice of the initial beam profile, besides the stable oscillating linear modes we have inevitably excited also the linearly unstable modes that eventually prevailed in our numerical example. However, the growthrate scales as $\sim\omega_{pp}/\gamma_u^2$ and in an experiment we can finely tune the initial beam profile so that such blowup or dispersion of the beam (i.e. the selffocussing or defocussing) is sufficiently delayed so that other physical effects enter the picture, including the two-stream instability, longitudinal (paraxial) variation, {and the nonlinear plasma wake force that can be responsible for the creation of different coherent nonlinear states such as 2D solitons.}

{It is important to note that the maximum value (in physical units) of the stationary solution (\ref{hole-like}), displayed in Fig. \ref{hole1}, is given by $n_{bc}\approx 4 N$. Using the normalizations Eq. (\ref{normalizations}), this readily yields the condition for the stationary state in the form $T_{bc}/m_0 c^2 = C\; n_{bc}/n_{p0}$ with $C\approx 0.63$. This expression coincides with that found earlier in the Vlasov description, using a macroscopic theory and the virial theorem, see Eq. (35) in \cite{2014EPJD...68..210F}, where the same functional dependence was obtained with the value $C\approx 1/2$. Such small discrepancy probably comes from the deviation of our solution (\ref{hole-like}) from the exact Gaussian, that was used in \cite{2014EPJD...68..210F} as the initial condition.}


\section{Conclusions}

In this paper we presented a self-consistent nonlinear hydrodynamic theory of the propagation of a long and thin relativistic electron beam through an unmagnetized and overdense plasma, in a configuration that is typical for the {plasma} wakefield acceleration scheme. The random component of the trajectories of the beam particles and of their velocities was modelled by an effective anisotropic temperature. It was demonstrated that in the presence of a finite velocity spread in the parallel direction and for not too long a beam, the beam dynamics could be approximated as a fully 3-D adiabatic expansion/compression. The resulting nonlinearity provided the localization of the beam in the transverse direction and produced a coherent stationary state, {even when the {effects of the   nonlinear plasma wake force are} small or absent}. The linear analysis revealed that a small perturbation of such coherent state could be unstable, resulting either in the defocussing or focussing of the beam when the number of the beam particles was initially larger or smaller than that of the stationary state. Numerical calculations demonstrated that the lifetime of the beam could be significantly extended by the appropriate fine tuning, so that the  transverse oscillations predicted earlier within the thermal wave model, could be observed. {Conversely, in a system that is predominantly two dimensional, this kind of thermodynamic nonlinearity vanishes and on the long timescale the system is governed by {the plasma wake force} responsible also for the betatron-like oscillations observed in the earlier works \cite{2014EPJD...68..210F,2014EPJD...68..271F,2014JPhCS.482a2014F,2016NIMPA.829..426A}.}

\begin{acknowledgements} This work was supported in part (D.J. and M.B.) by the MPNTR 171006 and NPRP 8-028-1-001 grants. D.J. acknowledges financial support from the INFN's fondo FAI and the hospitality of the Dipartimento di Fisica "Ettore Pancini", Universita di Napoli "Federico II", Italy.
\end{acknowledgements}

%\bibliography{Beam_evolution_2016}

\end{document}